# Biaxial strain tuning of the optical properties of single-layer transition metal dichalcogenides

Riccardo Frisenda*[1], Matthias Drüppel[2], Robert Schmidt[3], Steffen Michaelis de Vasconcellos[3], David Perez de Lara[1], Rudolf Bratschitsch[3], Michael Rohlfing*[2], Andres Castellanos-Gomez*[1]

[1] Instituto Madrileño de Estudios Avanzados en Nanociencia (IMDEA-nanociencia), Campus de Cantoblanco, E-28049 Madrid, Spain.

[2] Institute for Solid-state Theory, University of Münster, 48149 Münster, Germany.

[3] Institute of Physics and Center for Nanotechnology, University of Münster, 48149 Münster, Germany.

*E-mail: riccardo.frisenda@imdea.org; michael.rohlfing@uni-muenster.de; andres.castellanos@imdea.org.

**ABSTRACT:** Since their discovery single-layer semiconducting transition metal dichalcogenides have attracted much attention thanks to their outstanding optical and mechanical properties. Strain engineering in these two-dimensional materials aims to tune their bandgap energy and to modify their optoelectronic properties by the application of external strain. In this paper we demonstrate that biaxial strain, both tensile and compressive, can be applied and released in a timescale of a few seconds in a reproducible way on transition metal dichalcogenides monolayers deposited on polymeric substrates. We can control the amount of biaxial strain applied by letting the substrate expand or compress. To do this we change the substrate temperature and choose materials with a large thermal expansion coefficient. After the investigation of the substrate-dependent strain transfer, we performed micro-differential spectroscopy of four transition metal dichalcogenides monolayers ($MoS_2$, $MoSe_2$, $WS_2$, $WSe_2$) under the application of biaxial strain and measured their optical properties. For tensile strain we observe a redshift of the bandgap that reaches a value as large as 95 meV/% in the case of single-layer $WS_2$ deposited on polypropylene. The observed bandgap shifts as a function of substrate extension/compression follow the order $MoSe_2 < MoS_2 < WSe_2 < WS_2$. Theoretical calculations of these four materials under biaxial strain predict the same trend for the material-dependent rates of the shift and reproduce well the features observed in the measured reflectance spectra.

# INTRODUCTION

Strain engineering has been proposed as a prospective route to modify the electronic and optical properties of two-dimensional (2D) materials [1-4]. The interest in this topic is motivated by their resilience to mechanical deformations. These systems stand deformations of the order of 10% [5,6], while conventional 3D semiconductors break at quite moderate deformations of 0.5-1.5% [7]. Apart from the mechanical toughness, another key question of strain engineering of 2D materials is how they can be conveniently and reproducibly strained. While 3D systems are typically stressed by epitaxially growing them onto substrates with a certain lattice parameter mismatch, strain in 2D systems can be applied more directly by folding [8], stretching [9-14] and bending [15-25]. Moreover, strain in 2D materials can be easily varied continuously in time, helping to achieve a modulation of the electronic properties. Experiments on $MoS_2$ single-layer and few-layer flakes have already demonstrated that the optical band gap is tunable by 50 meV/% for uniaxial strain [15,16] and 100 meV/% for biaxial strain [14]. These results open the door to fabricate devices whose optical and electronic properties can be externally controlled by the application of strain [26].

Most of the strain engineering experiments to date, have been mainly focused on uniaxial strain under static conditions. However, time-dependent straining is desirable for many applications such as sensors, optical modulators or active optic devices. Here, we explore the use of biaxial strain to modulate the reflectivity of single layer transition metal dichalcogenides (TMDCs) on a timescale of a few seconds. We investigate the strain transfer from thermally expanded or compressed polydimethylsiloxane (PDMS) substrate to a $MoS_2$ monolayer. PDMS is a polymer commonly used in strain engineering experiments with 2D materials. However, we find that this substrate is not efficient in compressing or extending the 2D material on top. Switching to polypropylene (PP) substrates we find that biaxial strain can be applied reproducibly without slippage up to a maximum tensile biaxial strain of 1%. We investigate the effects of strain on the optical properties of single-layers $MoS_2$, $MoSe_2$, $WS_2$ and $WSe_2$. To our knowledge this is the first experimental investigation of $MoSe_2$, $WS_2$ and $WSe_2$ under the application of biaxial strain. For increasing tensile strain a redshift of the optical band gap of these 2D TMDCs is observed, that reaches, in the case of $WS_2$, the large value of 95 meV for a substrate strain of only 1%. The observed bandgap shifts as a function of substrate extension/compression follow the order
$MoSe_2 < MoS_2 < WSe_2 < WS_2$, i.e. with $WS_2$ providing the largest bandgap tunability and $MoSe_2$ the lowest. Using the thermal expansion mismatch between a 2D material and a substrate is a simple but powerful way to achieve biaxial expansion or compression of the 2D material, which is technically more difficult to achieve than uniaxial strain. This method can be readily applied to other 2D materials and be used to vary the strain in real time.

In order to apply biaxial strain to single-layer TMDCs we change the temperature of the substrate with a Peltier heater/cooler and exploit the large mismatch between the thermal expansion coefficients of the substrate and the TMDC flake deposited on top, similarly to previous work by part of the authors [12]. In this work, however, it was not possible to directly determine the substrate expansion and the biaxial strain was limited only to tensile strain. The selected substrate could not effectively transfer the strain to the 2D layer because of its low Young's modulus and only static strain was studied [11,27]. Here, we present a simple method to accurately calibrate the substrate expansion and we extend the straining method also to compressive strain. By analyzing the nature of the strain transfer mechanism we find a substrate that optimizes it and we study time-dependent strain.

**RESULTS**

Figure 1a shows a sketch of the thermal expansion calibration method. In the calibration procedure we measure the distance between periodical features, patterned on the surface of the polymeric substrate by recording optical images of the substrate while changing the temperature with a Peltier element (10 - 110 °C). Figure 1b displays two fragments of optical images of a PDMS substrate with periodic holes (diameter 3 μm and pitch 4 μm) taken at a temperature of 30 °C and 110 °C. Thanks to the presence of the array of holes the thermal expansion of the PDMS substrate is readily visualized when comparing the two images.

We analyze the spacing between holes at different temperatures by studying the autocorrelation function of the intensity of the images, which provides a powerful tool to extract periodical features as explained in Section S1 of the Supplementary Information. For each temperature $T$ we extract the distance $L(T)$ between adjacent holes. The change in distance between holes $\Delta L$ is related to the change in temperature $T$ according to:

$$\frac{L(T)-L(T=25\ °C)}{L(T=25\ °C)} = \frac{\Delta L}{L} = \alpha_{Sub} \cdot (T - 25\ °C),\qquad 1$$

where $\alpha_{Sub}$ is the thermal expansion coefficient of the material. Figure 1c shows the percentage change in distance as a function of the temperature of the PDMS substrate. The observed dependence is linear and the linear thermal expansion coefficient can be directly extracted from this measurement, yielding a value $\alpha_{PDMS}$ = (3.4 ± 0.3) · $10^{-4}$ °$C^{-1}$, in good agreement with the value reported in literature [28]. One can easily determine the amount of biaxial expansion/compression $\varepsilon_{Sub}$ of the substrate directly from the temperature $\varepsilon_{Sub} = \alpha_{Sub} \cdot (T - 25\ °C)$. The thermal expansion of this substrate is roughly 50 times larger than that of TMDCs [5,29]. Therefore, a change in temperature of the substrate is expected to yield a biaxial strain of the TMDCs deposited on the substrate.

We first studied the effect of strain on mechanically exfoliated flakes of monolayer $MoS_2$ deposited on PDMS, a substrate that is commonly used in strain engineering experiments with 2D materials. The flakes have been deposited onto the PDMS substrates by mechanical exfoliation with Nitto SPV 224 tape. Subsequent deterministic placement [30] allows one to transfer the flakes from the PDMS to PP and PC substrates. The thickness of the flakes is determined by a combination of quantitative optical analysis, differential reflectance spectroscopy and Raman spectroscopy. Figure 2a shows an optical image of a flake of $MoS_2$ deposited on a PDMS substrate with different numbers of layers of $MoS_2$ showing a different optical contrast. Figure 2b displays the Raman signal, measured at the point indicated in Figure 1a, which shows two prominent maxima, which are well fitted to Lorentzian functions and attributed to $A_{1g}$ and $E^1_{2g}$ vibrations of $MoS_2$ [31,32]. We find a frequency difference $\Delta f$ = (19.6 ± 0.2) $cm^{-1}$ between these vibrations, which confirms that the dark region in the optical image is a monolayer of $MoS_2$, since it is known that $\Delta f$ increases with the number of layers as shown in the inset of Figure 2b.

To study the optical properties of the monolayer as a function of biaxial strain we use a micro-reflectance setup in which we illuminate a small area (diameter ≈ 60 µm) of the sample with a halogen white light source, which is collimated through a small diaphragm and focused with the microscope lens (see Supplementary Information Fig. SI5). Thanks to a fiber optic (core diameter 105 µm) attached to the trinocular, we collect only the light reflected by a small area of the sample (diameter 2.1 µm, see Supplementary Information Fig. SI6). The dimensions of this probed area are much smaller than a typical single-layer region that measures at least 10 x 10 µm. The differential reflectance signal of an ultra-thin film adsorbed on a surface is directly proportional to the absorbance of the film [33,34], see section 3 of the Supplementary Information. Figure 2c shows differential reflectance spectra acquired on the single-layer region of the $MoS_2$ flakes. The differential reflectance spectra of single-layer $MoS_2$ increase for energies larger than the bandgap (optical band gap: 1.9 eV). On top of a broad background one finds two prominent maxima centered at 1.90 eV and 2.05 eV, which correspond to the A and B excitons. Photons are absorbed due to the direct transitions at the K point of the Brillouin zone of monolayer $MoS_2$ [35-38].

The differential reflectance spectra have been recorded at room temperature (25 °C) and at 55 °C. This temperature difference causes an expansion of the substrate of approximately 1%. An inspection of the spectra reveals that both the A and B resonances display a redshift of 13 meV for this substrate expansion. To extract the energy of the excitons from Figure 2c, we fit the peaks present in the differential reflectance spectra with Lorentzian functions. In Figure 2d we display the energy shift of the A and B excitons, with respect to the room temperature values, as a function of the thermal expansion of the substrate. The data indicate a linear dependence on the substrate expansion both for tensile and for compressive strain. We find a gauge factor, *i.e.* the rate of shift in energy of a spectral feature as a function of the substrate percentage expansion/compression, of approximately -13 meV/% (corresponding to -0.44 meV/°C, see Table 1). To test the reproducibility and rule out slippage we studied the differential reflectance of single-layer $MoS_2$ during consecutive cycles of warming/cooling of the PDMS substrate (see Supplementary Information Fig. SI16, additional measurements on polycarbonate are discussed in the in Fig. SI12).

To investigate the intrinsic effect of the temperature on the optical spectrum of a single layer $MoS_2$ flake we repeated the warming/cooling experiment using glass as a substrate. Glass has a small thermal expansion coefficient ($\alpha_{Glass}$ = 0.04 · $10^{-4}$ °$C^{-1}$), which is approximately two orders of magnitude smaller than the coefficients of polymeric substrates, resulting in an expected maximum strain attainable of only 0.04% when heating from 10 °C to 110 °C. Figure SI11 of the Supplementary Information shows the differential reflectance spectra of single-layer $MoS_2$ deposited on glass for different temperatures. The A and B excitons redshift for increasing temperature. From a linear fit we find the intrinsic thermal dependence of the energy of the A and B excitonic peak of monolayer $MoS_2$ equal to -0.34 meV/°C and -0.42 meV/°C, respectively. These values are in good agreement with previous measurements of $MoS_2$ on $SiO_2$ and of single-layer $MoSe_2$ on $SiO_2$ [35,39] and are attributed to the redshift of the bandgap energy caused by the thermal expansion of the lattices of these single-layer TMDCs. After subtraction of this intrinsic thermal shift from the measured gauge factor of $MoS_2$ on PDMS, the energy shift of the excitons induced by straining the polymer substrate is between 0 and -2 meV/%. In contrast, our ab-initio calculations of the absorption spectra of $MoS_2$ upon biaxial strain show a much more pronounced shift. We perform density functional theory (DFT) calculations in the local density approximations, followed by a GW step within the LDA+GdW approximation [40], to then solve the Bethe-Salpeter equation (BSE) to access absorption spectra. Figure 3 displays our results for a single-layer $MoS_2$ in absence of external strain and with 1% of biaxial tensile and compressive strain. We find a linear dependence of both the A and B excitons. The gauge factors for the quasiparticle gap and the A and B excitons are compared in Table 2. An extended table including the quasiparticle gaps and energetic positions of the A and B excitons can be found in the Supplementary Information (Table SI2). The difference between the calculated gauge factors and those measured using PDMS substrates is a factor of 100 smaller. The fact that the redshift observed for $MoS_2$/PDMS is comparable in magnitude to the one observed for $MoS_2$/glass, even if the strain of PDMS is a factor 50 larger than the strain attainable in glass, together with the large discrepancy between the theoretical and the experimental gauge factors indicate that most of the strain present in the PDMS substrate is not transferred to the adsorbed $MoS_2$ flake.

According to previous studies on graphene [27] the efficiency of a substrate to transfer strain to a flake depends on the Young's modulus of the substrate $E_{Sub}$. The expected maximum strain $\varepsilon$ induced in the monolayer TMDCs flakes is typically a fraction of the substrate strain $\varepsilon_{Sub}$ and can be written as:

$$\varepsilon = g \cdot \varepsilon_{Sub},  \quad\quad 2$$

with $g$ being a dimensionless parameter that depends on $E_{Sub}$ and on the lateral size of the monolayer and has values between 0 and 1 [11]. This is supported by finite element simulations (shown in Figs. SI17 and SI18 of the Supplementary Information) in which we perform an axisymmetric simulation of a MoS$_2$ flake (diameter 10 μm, height 0.7 nm) on top of a substrate (diameter 500 μm, height 100 μm). We calculate the amount of strain transferred from the expanded substrate to the MoS$_2$ flake ($E_{MoS_2}$ ≈ 350 GPa [11]) as a function of the substrate's Young's modulus and we find for substrates with a Young's modulus comparable to that of PDMS ($E_{Sub}$ ≈ 500 kPa) that the strain transferred is on the order of 1% ($g$ = 0.01). To transfer a larger amount of strain one should use substrates with Young's moduli larger than 500 MPa. Among the different possible materials we choose PP as a substrate because of the good trade-off between its thermal expansion coefficient ($\alpha_{PP}$ ≈ 1.35 · 10$^{-4}$ °C$^{-1}$) and its Young's modulus ($E_{PP}$ ≈ 1.5 GPa), which according to the finite element simulation give a strain transfer efficiency $g$ = 0.75 for a flake of MoS$_2$.

We performed differential reflectance measurements as a function of substrate strain on single-layer flakes of MoS$_2$, MoSe$_2$, WS$_2$ and WSe$_2$ deposited on PP (see Fig. SI13 of the Supplementary Information for similar measurements performed on PDMS). Figure 4 displays the differential reflectance spectra of the four TMDCs single-layer flakes recorded at zero substrate expansion and at 0.9% of biaxial expansion and at 0.1% of biaxial compression. All the spectra show maxima attributed to excitons in the materials [36,37] on top of a broad background. An accurate theoretical description of the features present in the optical absorption spectra can be found in Ref. [41]. At zero strain and energy lower than 2.2 eV, the Mo based TMDCs show two peaks, already discussed in the case of MoS$_2$, labelled A and B, whose separation is essentially equal to the spin-orbit splitting of the valence band. In the W-based TMDCs, the A peak is still evident, while the B exciton is less prominent, giving rise to a shoulder in WS$_2$ around 2.4 eV and to a peak at 2.1 eV in WSe$_2$. The larger spin-orbit splitting, due to the heavier W atoms compared to Mo atoms, induces a larger separation of the A and B features in the differential reflectance spectra of W-based TMDCs. The pronounced broad peak at energies above 2.5 eV present in MoS$_2$, MoSe$_2$ and WS$_2$ (C exciton) has been interpreted as coming from nearly-degenerate exciton states, located in regions of the Brillouin zone where the valence and conduction bands are nested. The different and more complicate physical origin of the peak C in respect to the A and B excitonic peaks makes a direct comparison of the behavior under strain difficult.

The A and B features shift to lower energies when the substrate expands and to higher energies for substrate compression. We extract the position of the A and B features as a function of substrate temperature for the four TMDCs and plot the results in Figure 4. A linear fit to the data allows determining the gauge factors that are listed in Table 2 for each material. The magnitude of the shifts induced by the application of strain follow the order WS$_2$ > WSe$_2$ > MoS$_2$ > MoSe$_2$. Thus, given the same chalcogenide atom (S or Se), W atoms induce a larger gauge factor than Mo atoms, due to their more diffuse $d$ orbitals. Conversely, given the same metal, the lighter S atoms induce a larger shift than Se atoms. These effects and the trend observed are well reproduced by the calculations reported in Figure 3 and in Figure SI20 of the Supplementary Information.

**DISCUSSION**

Our theoretically obtained gauge factors are systematically higher than our experimental values. In the experiment, as already discussed, only a part of the strain is transferred from the substrate to the monolayer. Therefore, the calculated values are an upper bound for the experimental gauge factors. The ordering of the gauge factors on the other hand for the different materials ($WS_2$ > $WSe_2$ > $MoS_2$ > $MoSe_2$) is perfectly reproduced by our calculations. The magnitudes and ordering of the theoretical exciton gauge factors closely follow the gauge factors of the quasiparticle gap, as can be seen in Table 2, demonstrating that the underlying electronic structure already dictates the ordering. We find that the shift in the gaps under strain mainly stems from the shift of the conduction band minima (CBM). In the CBM the shifts (with respect to the vacuum energies) are three times as high as in the valence band minima (VBM) for all four TMDCs. The magnitude of the gauge factor ultimately measures how strong the electronic structure of the monolayer reacts to structural change. Strongly overlapping wave functions between atoms would lead to a larger gauge factor. The band width in the conduction band can be taken as a good indicator for the magnitude of the inter-atomic orbital overlap. Comparing the band width of the lowest conduction band around the K point (within a distance of 0.2 π/a from K, the area of momentum space which is mainly responsible for the A and B excitons) we find the ordering $WS_2$ (0.30 eV) > $WSe_2$ (0.26 eV) > $MoS_2$ (0.20 eV) > $MoSe_2$ (0.18 eV) that follows the same ordering of the exciton gauge factors discussed previously. In conclusion, this suggests that the magnitude of the gauge factor, which mainly stems from the change in the CBM with strain, is largest for a strong overlap between atoms, which can be estimated through the band width around the K point. We notice that for the values of biaxial strain investigated, the four single-layer TMDCs remain direct gap semiconductors [4,13,18,39,42].

The strain-induced shift of the excitonic peaks in the differential reflection spectra of single-layer TMDCs is particularly interesting for applications as optical modulators. By controlling the temperature of the polymeric substrate, we can easily tune the position of the excitonic peaks and achieve a 10% modulation of the reflection for certain wavelengths in a time-scale of a few seconds. We record the differential reflectance of a $MoS_2$ single-layer flakes deposited on PP while cycling the temperature between 30 °C and 75 °C, applying a square wave modulated voltage to the Peltier heater. Figure 5a displays the differential reflectance at an energy of 1.91 eV, corresponding to a wavelength of 648 nm, recorded as a function of time with a resolution of 300 ms. The reflectance has a distorted square wave profile with rise/fall characteristic times of 10 seconds with a modulation of the intensity of 8% at 648 nm. This timescale is limited by the heat transfer between the Peltier heater and the polymeric substrate and could be improved by using local micro-heaters. Figure 5b displays the position of the A exciton peak as a function of time. The reproducibility in the shift and the transfer rate of the strain are excellent. The observed modulation in the signal of the order of 10% (at specific wavelengths) is noteworthy especially considering the atomic thickness (< 1 nm) of these single-layer TMDCs. Modulations up to 25% can be reached thanks to the excitons that dominate the dielectric function of single-layer TMDCs materials which are present at room temperature.

In conclusion, we have exploited the large thermal expansion coefficient of a polymer substrate to apply large biaxial tensile strain on single-layer flakes of four transition metal dichalcogenides. By recording the exciton-dominated differential reflectance spectra of these materials as a function of external strain we monitored the change in bandgap induced by the strain. We observe that the magnitude of the induced energy shift is the largest with $WS_2$ flakes where it reaches 95 meV/%. The shift follows the order $MoSe_2$ < $MoS_2$ < $WSe_2$ < $WS_2$. Theoretical calculations reproduce well the experimental results and the observed trend in the various materials for the dependence of the bandgap on the external strain. The large shift induced in the bandgap by biaxial strain and the rapidity for the transfer of strain open the possibility for the use of 2D TMDCs as electro-optical modulators or strain sensors.

## METHODS

*Differential reflectance measurements*

White light from a halogen lamp irradiates the sample after passing through a small diaphragm and results in an illumination spot of approximately 50 μm at the sample surface. The light reflected from the sample is collected with an optical fiber (105 μm core diameter) and feed to a spectrometer. The fiber is used to collect only the light reflected from a few μm area of the sample located approximately in the center of the illumination spot. The differential reflectance spectrum of a TMDC flake is calculated by subtracting from the reflectance spectrum collected on top of the flake the same spectrum collected on the substrate, and normalizing the result by the flake spectrum.

*Ab-initio calculations*

We performed ab-initio density functional theory (DFT) calculation in the local density approximation (LDA), from which we obtain the energetically optimized structure. The DFT wave functions and energies are then used as input for a subsequent GW calculation within the LDA+GdW approximation [36], in which the dielectric screening properties are described by an atom-resolved model function based on the random-phase approximation [36]. In the last step, the Bethe-Salpeter equation (BSE) is set up from the quasiparticle band structure. For the given TMDC monolayers, excitons are composed from four valence and six conduction bands and a mesh of 30 x 30 x 1 k points from the first Brillouine zone. To simulate biaxial tensile (compressive) strain, the lattice constant is increased (decreased) fully relaxing the structure for each applied strain. All further numerical details can be found in the Supplementary Information. This approach has already been successfully used to describe uniaxial strain in a $WSe_2$ monolayer [19].

## ACKNOWLEDGEMENTS

We thank Dr. Isabel Rodriguez for the nanoimprinted PDMS substrates and Dr. Emilio Pérez Alvarez and Alejandro López Moreno for their help with the Raman measurements. We acknowledge financial support from the European Commission, the MINECO, the Comunidad de Madrid and from the Netherlands Organisation for Scientific Research (NWO). M.D. thanks the Studienstiftung des deutschen Volkes for support via a PhD fellowship. The authors gratefully acknowledge the computing time granted by the John von Neumann Institute for Computing (NIC) and provided on the supercomputer JURECA at Jülich Supercomputing Centre (JSC).

## COMPETING INTERESTS

The authors declare no competing financial interests.

## CONTRIBUTIONS

R.F. and A.C.G. carried out the strain experiments. R.F analyzed the experimental data. M.D and M.R. performed the ab-initio calculations. A.C.G. and D.PdL. designed the experiment. R.S., S.M.dV. and R.B. performed the finite element simulations. All the authors discussed the results and contributed to the writing of the manuscript.

**FUNDING**

A.C.G. European Commission under the Graphene Flagship: contract CNECTICT-604391

A.C.G. MINECO: Ramón y Cajal 2014 program RYC-2014- 01406

A.C.G. MINECO: program MAT2014-58399-JIN

A.C.G. Comunidad de Madrid: MAD2D-CM Program (S2013/MIT-3007)

R.F. Netherlands Organisation for Scientific Research (NWO): Rubicon 680-50-1515

D.P.dL. MINECO: program FIS2015-67367-C2-1-p

**FIGURES**

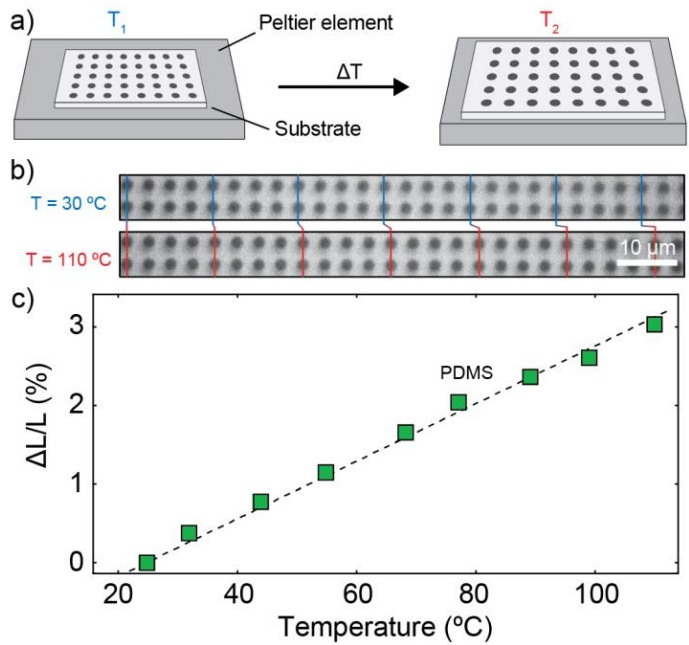

**Figure 1**: a) Schematic of the temperature-dependent experiment. Substrate heating (cooling) causes the thermal expansion (contraction) of the substrate that induces biaxial strain on the flake pre-deposited on top. Control over the temperature is achieved with a Peltier heating/cooling element. b) Optical microscope image (in gray scale) of a PDMS substrate with periodic holes taken at two different temperatures (30 and 110 °C). Notice the expanded substrate at higher temperature. c) Average percentage increase of the distance between holes as a function of temperature. The dashed line is a linear fit to the data. The slope yields the expansion coefficient of PDMS.

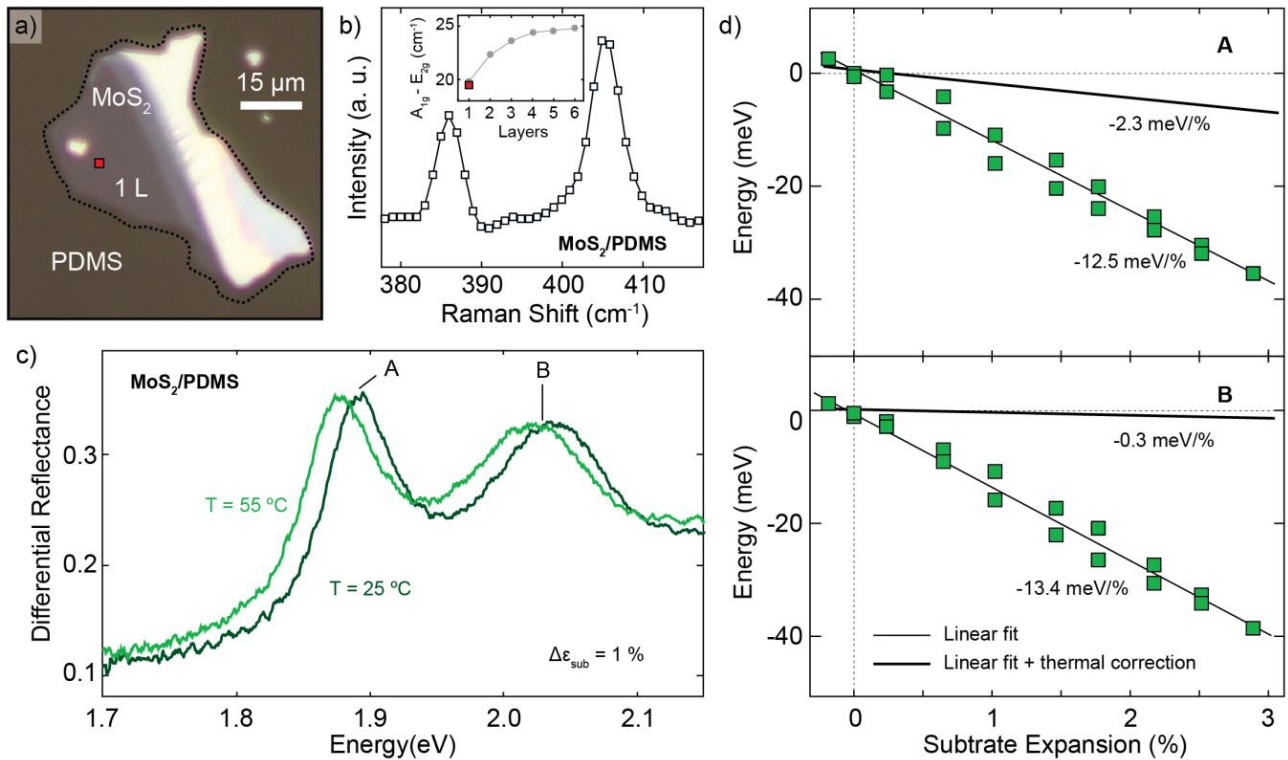

**Figure 2**: a) Optical image of a MoS$_2$ flake deposited on a PDMS substrate; the darker violet region is single-layer MoS$_2$. The contour of the flake is outlined by a black dashed line. b) Raman spectrum of the MoS$_2$ flake taken at the position of the red square in panel (a). The inset shows the dependence of the difference in Raman shift of the modes A$_{1g}$ and E$_{2g}$ as a function of the number of layers, the red square indicates the value found in this study. c) Differential reflectance measured on the single-layer region of the MoS$_2$ flake for zero substrate expansion (light colored curve) and at 1 % of substrate expansion (dark curve). d) Energy of the excitonic peaks A, B in panel (c), extracted from a multi-peak fit as a function of the substrate expansion. The thin lines represent a linear fit to the data, while the thick line represents the same trend after subtraction of the intrinsic thermal dependence contribution.

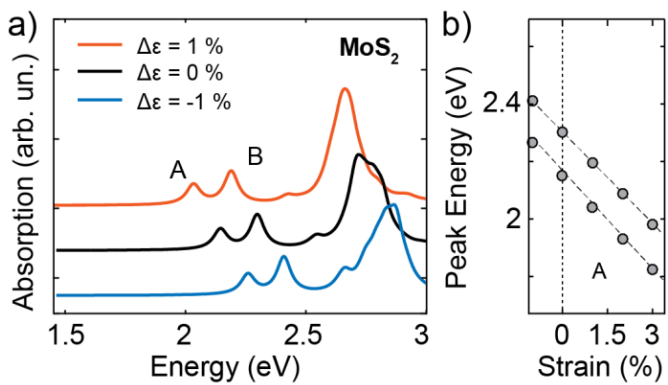

**Figure 3**: Calculated BSE absorption spectra under biaxial strain of single-layer MoS$_2$ (left). An artificial broadening of 0.035 eV is used in the BSE calculations and the spectra are vertically shifted for improved visibility. The energies of the A and B excitons are extracted from the ab-initio calculations (right).

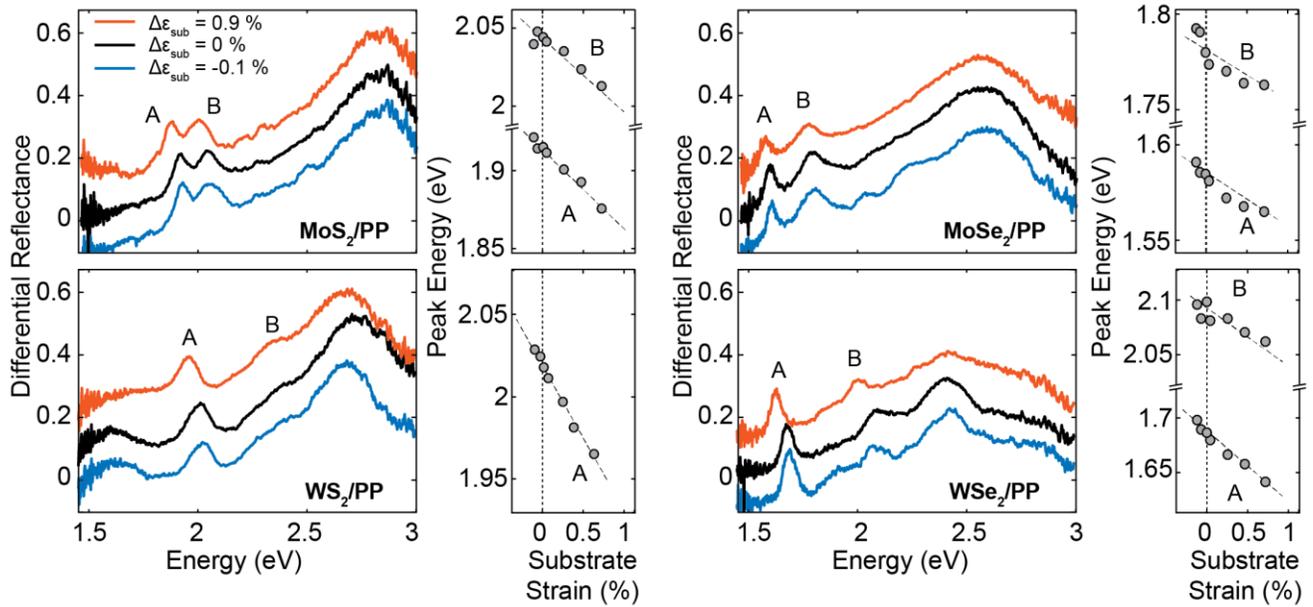

**Figure 4**: Differential reflectance spectra of single-layer flakes of the four TMDCs deposited on PP. The spectra have been measured as a function of the substrate expansion (red curves) and contraction (black curves) and are vertically shifted for clarity. Energy of the excitonic peaks (labelled A and B) as a function of substrate strain extracted from the differential reflectance spectra.

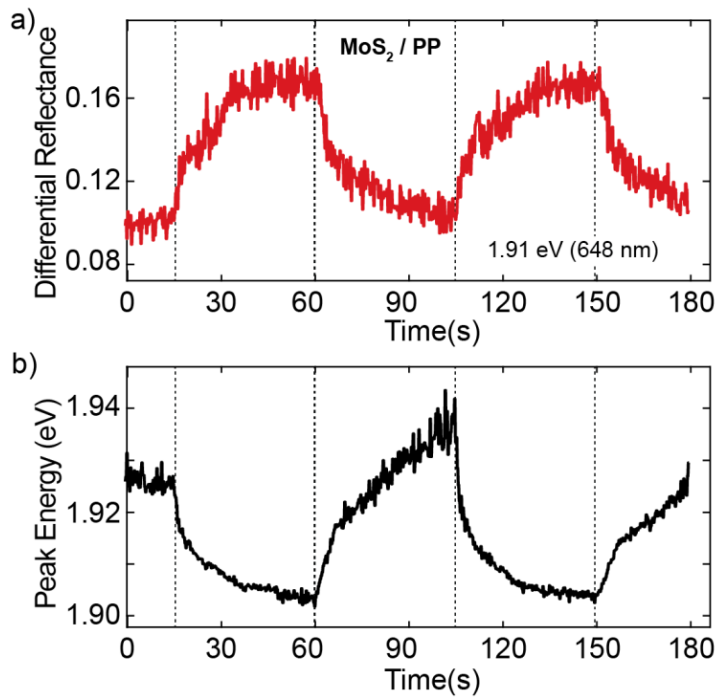

**Figure 5**: a) Differential reflectance measured at a wavelength of 648 nm on single-layer $MoS_2$ deposited on PP as a function of time with periodical heating and cooling cycles of the substrates from 30 °C to 75 °C. Each differential reflectance spectrum has been integrated for 300 ms. b) Energy of exciton A as a function of time.

**TABLES**

| Exciton | MoS$_2$/PDMS | MoS$_2$/PP |
|---|---|---|
| A | -12.5 meV/% <br><br> (-2.2 meV/%) | -51.1 meV/% <br><br> (-25.4 meV/%) |
| B | -13.4 meV/% <br><br> (-1.0 meV/%) | -48.7 meV/% <br><br> (-17.8 meV/%) |

**Table 1**: Gauge factor for excitons A and B of MoS$_2$ single-layer flakes deposited on PDMS and PP substrates extracted from applying biaxial strain on the different two-dimensional flakes. The numbers between brackets are the gauge factor with the thermal component removed.

| Exciton | MoS$_2$/PP | MoSe$_2$/PP | WS$_2$/PP | WSe$_2$/PP |
|---|---|---|---|---|
| A (experiment) | -51 meV/% | -33 meV/% | -94 meV/% | -63 meV/% |
| B (experiment) | -49 meV/% | -30 meV/% | | -43 meV/% |
| A (theory) | -110 meV/% | -90 meV/% | -151 meV/% | -134 meV/% |
| B (theory) | -107 meV/% | -89 meV/% | -130 meV/% | -111 meV/% |
| Quasiparticle band gap (theory) | -134 meV/% | -115 meV/% | -156 meV/% | -141 meV/% |

**Table 2**: Gauge factor for excitons A and B; the experimental values are extracted from applying biaxial strain on the different two-dimensional flakes. The theoretical values stem from the calculated absorption spectra (BSE). Also included are the gauge factors of the quasiparticle gap, i.e. the VBM to CBM direct band gap at the K point.

**Supplementary Information**

**Section 1 – Thermal expansion of polymeric substrates**

To calibrate the expansion of the substrates, we use a microscope to take pictures of µm-sized periodical features patterned on top of the substrate, while heating or cooling the substrate with a Peltier element. Figure SI1a displays two optical images of the same PDMS substrate (2 cm x 2 cm, height 1 mm) with µm-sized periodical holes, recorded at two different temperatures. The expansion of the PDMS at higher temperature can be directly visualized thanks to the larger relative separation of the holes in the image. Figure SI1b displays the autocorrelation of the line profile of the two images taken along the rows indicated in panel a, which is a valid tool to extract the periodicity of a signal even in presence of noise. The curve taken at 110 °C has indeed a larger period than the curve taken at 32 °C. The large expansion of the substrate can also be discerned by the naked eye as can be seen in Figure SI1c, where we draw the position of the holes in the two images of panel to show the uniform expansion of the substrate.

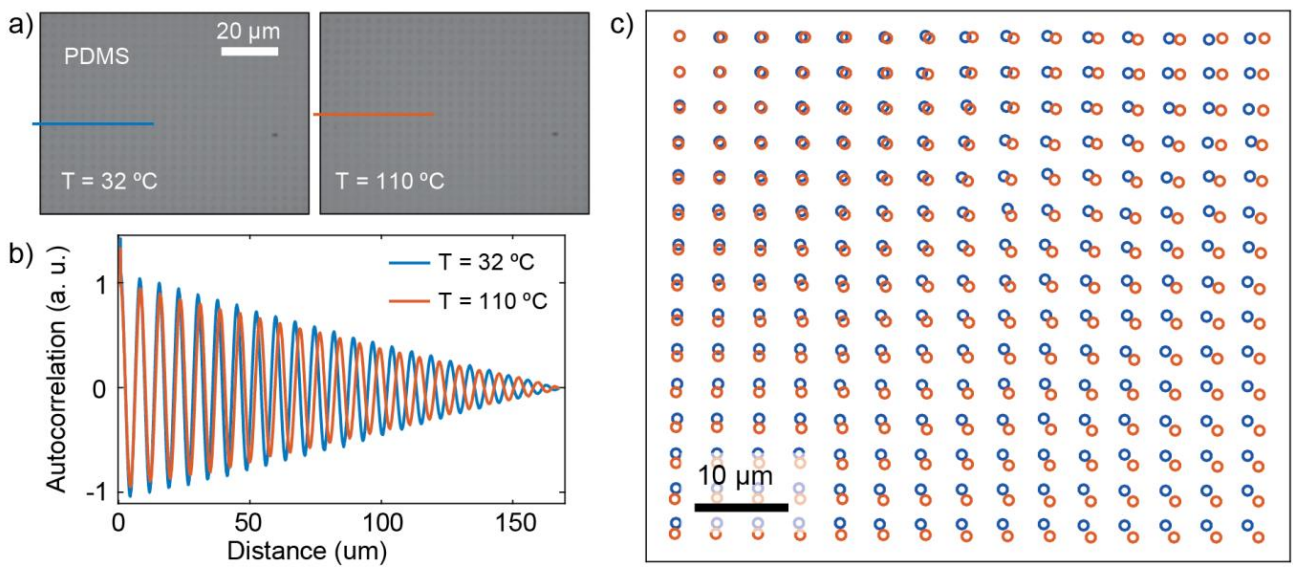

**Figure SI1**: a) Optical microscope image (in gray scale) of a PDMS substrate with periodic holes, taken at a temperature of 32 °C (left) and 110 °C (right). b) Autocorrelation function of the intensity of the line-profiles taken at the positions indicated by the lines in (a). c) Spatial map of the centers of the holes extracted from the two images in (a), where blue circles correspond to the data at 32 °C and red circles to the data at 110 °C.

From the position of the holes at the different temperatures one can construct a displacement map, which shows the change in position of each hole due to the expansion. Figure SI2a displays such a map where the displacement of each hole from its initial position is represented with an arrow. By dividing the intensity of each displacement point by the distance of that point from a common origin (taken at the center of the uppermost left hole), we have built a spatial strain map, shown in Figure SI2b, where strain is represented in color. Inspecting the color map, one observes a uniform green color, corresponding to 2.4% of strain that indicates a uniform biaxial expansion of the substrate.

| Material | $\alpha_L$ ($10^{-6} \cdot 1/K$) | Young's Modulus (GPa) |
|---|---|---|
| PDMS | 340 ± 15 | 0.00036 - 0.00087 |
| PP | 136 ± 15 | 1.5 – 2.0 |

**Table SI1**: Material properties of the substrates PDMS and PP. $\alpha_L$ has been found experimentally from the graph in Figure SI4 and the Young´s modulus is taken from reference 26 of the main text.

To extract the thermal expansion coefficient α of each substrate material reliably we perform a statistical analysis of the separation between holes as shown in Fig. SI3, where we show the process for the substrate pictured in panel a. Thanks to a Matlab algorithm we extract the line profile along each row and column of pixels in the image, Fig. SI3a, and calculate the autocorrelation function of each of these line profiles. We then search for consecutive peaks in the autocorrelation function, see Fig. SI3b, and build a histogram from all the extracted separation values, Fig. SI3c. We then fit a Gaussian function to the peaked histogram, whose center and variance permit to estimate the average separation, $L(T)$, between consecutive peaks at each temperature $T$. Finally α can be found from the plot of the percentage increase in length, $\frac{L(T)-L(T=25\,°C)}{L(T=25\,°C)}$, as a function of the temperature. Figure SI4 shows such the percentage increase in length for PDMS, PP and polycarbonate (PC) extracted from optical images. The dashed lines in the figure are linear fits to the data and the slope of each of these lines is the thermal expansion coefficient of the material. Table SI1 contains the linear thermal expansion coefficient (extracted from the experiment) and the Young's modulus (literature) of PDMS and PP.

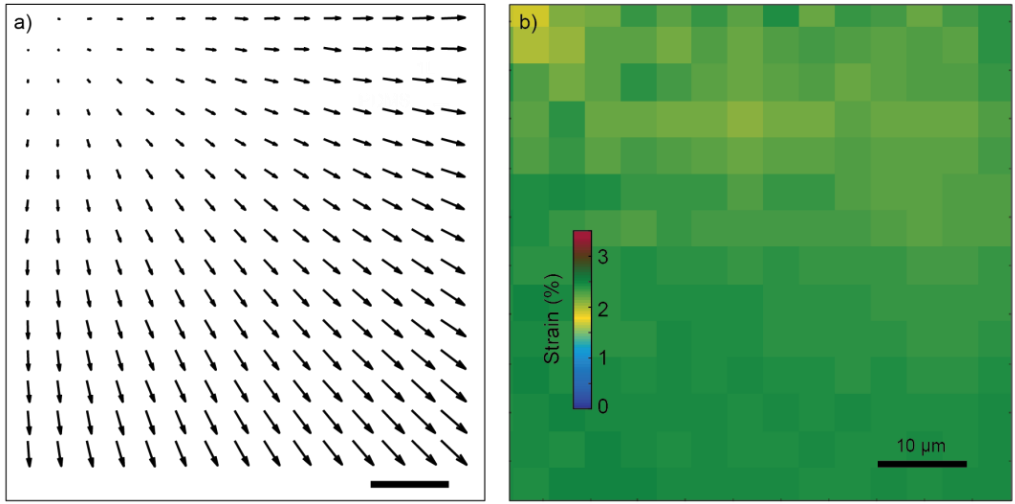

**Figure SI2**: a) Displacement map of substrate shown in Fig.SI1. b) Two-dimensional strain map extracted from the image in (a).

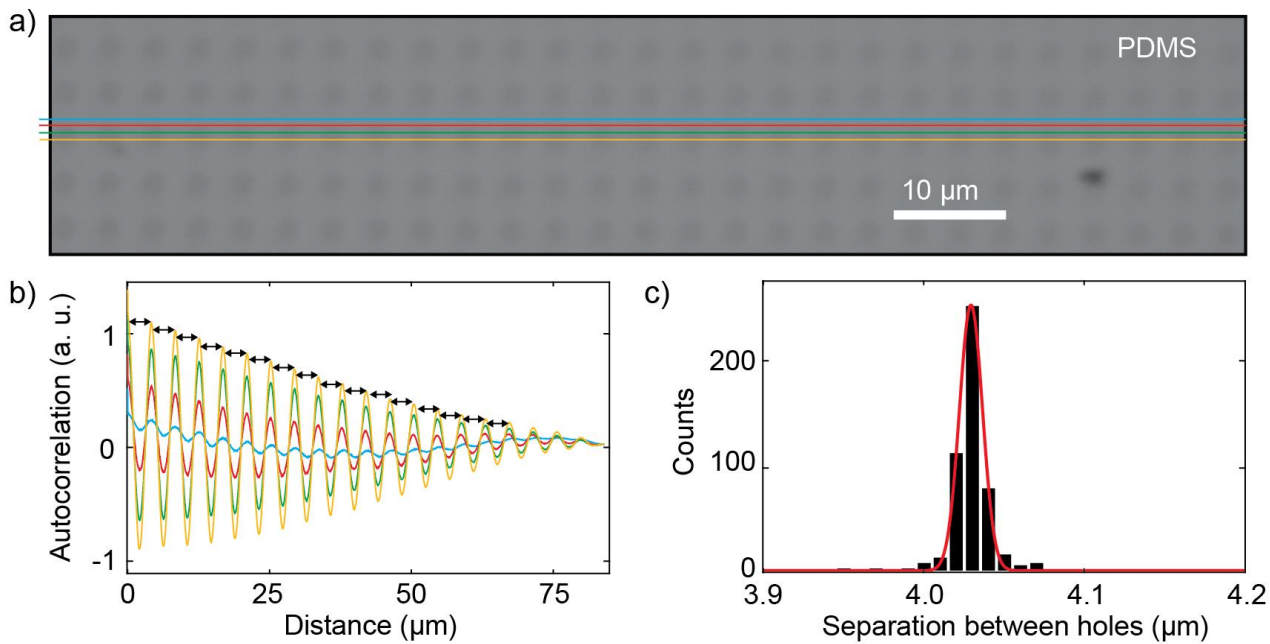

**Figure SI3**: a) Optical microscope image (in gray scale) of a PDMS substrate with periodic holes. b) Autocorrelation function of the intensity of the line-profiles taken at the positions indicated by the lines in (a). The black arrows represent the distance between consecutive maxima of the function. c) Histogram of the separation of consecutive peaks in the autocorrelation functions. The red line is a fit to a Gaussian peak.

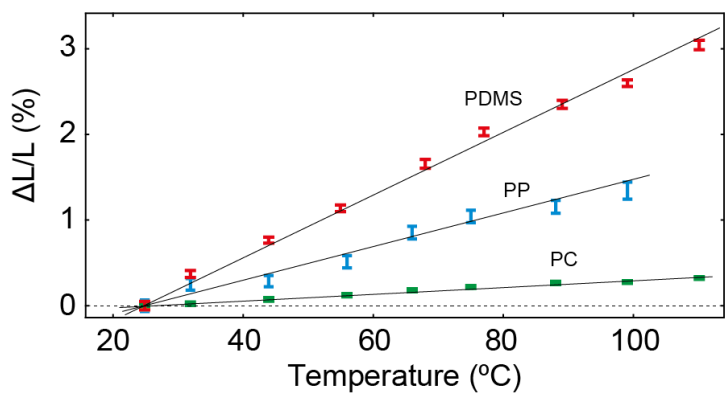

**Figure SI4**: Average percentage increase of the distance between periodic holes as a function of temperature. The solid lines are linear fit to the data; the slope of each line gives the expansion coefficient of PDMS, PP and PC.

### Section 2 – Differential reflectance setup

The differential reflectance spectra have been recorded in a set-up based on a commercial Motic microscope. Figure SI5 displays a schematic drawing. Briefly, a beam of white light, generated with a halogen lamp, is shined perpendicularly on the sample after passing through a small diaphragm which gives an illumination spot of approximately 50 µm at the height of the sample surface. The light reflected from the sample is then collected with a fiber optic of 105 μm of core diameter and feed to a Thorlabs spectrometer. The fiber optic is used to collect only the light reflected perpendicularly from a few-μm area of the sample located approximately in the center of the illumination spot. FigureSI6 shows an optical picture of a flake with visible the 50 μm area illuminated by the white light and the area cover by the fiber optic, visible as a bright spot in the middle of the image. The intensity line profiles taken along the x and y directions of Fig.SI6b reveal that the size of the collecting area is approximately 2 μm x 2 μm and that the spot is uniform in both directions.

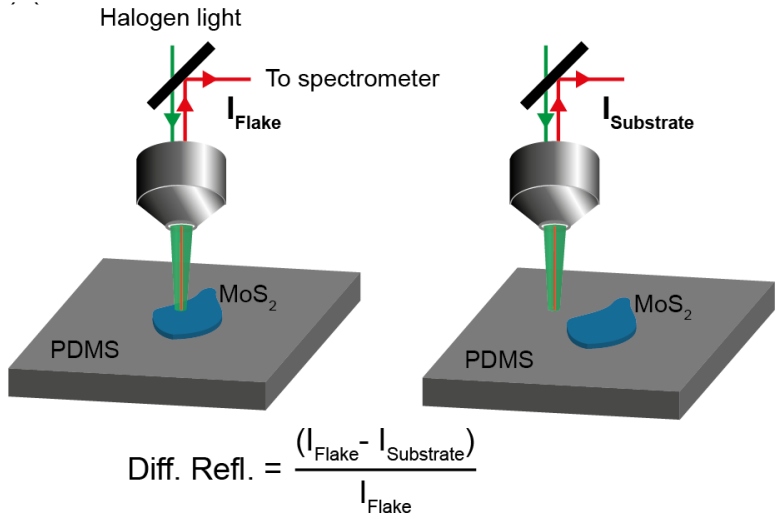

Figure SI5: Schematic drawing of the micro-reflectance set-up and of a micro-differential reflectance experiment.

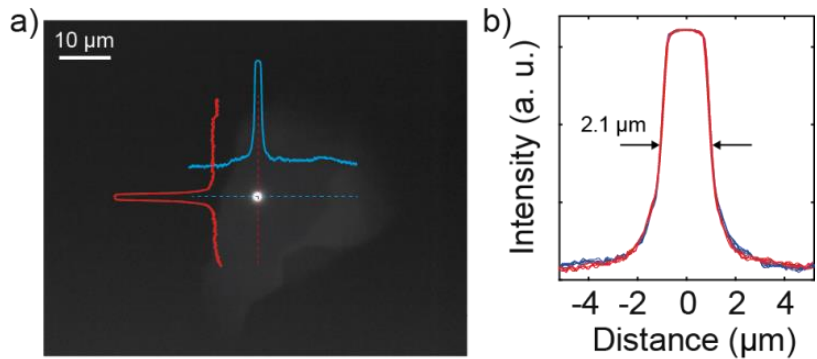

Figure SI6: a) Optical microscope image (in gray scale) of a MoS$_2$ flake deposited on the PDMS substrate. The bright spot in the center represents the spot size area of the differential reflectance measurement. b) Line profiles along the vertical (red) and horizontal (blue) directions of the spot size taken along the dotted lines in (a).

Section 3 – Optical properties of polymeric and glass substrates and differential reflectance of a thin film

The single-layer transition metal dichalcogenides flakes were deposited on different substrates (PDMS, PP, PC and glass). The substrates are transparent to visible light and we characterized them through transmittance measurements schematically depicted in Fig. SI5a. Figure SI5b displays the energy-resolved transmittance of the four substrates. All of the substrates have a transmittance that do not present features in the probed energy range. Since the substrates have different thicknesses we calculate the absorptivity from each transmittance curve, using the Beer-Lambert law. Figure SI5c collects the four absorptivity curves.

The differential reflectance (D.R.) spectrum of a thin film adsorbed on top of a transparent substrate measured at normal incidence is directly proportional to the absorption properties of the film, according to the following relation [1, 2]:

$$D.R. = \frac{4}{n_S^2 - 1} \cdot n \cdot \alpha(\lambda),  \quad\quad\quad 1$$

where $\alpha(\lambda)$, $n$, and $n_s$ represent the absorption of the film, refractive index of the film (n = 4 in the case of single-layer MoS$_2$), and refractive index of the substrate, respectively. Figure SI8a displays the energy resolved refractive index of PP and PDMS extracted from references [3,4]. Figures SI8b and SI8c show the absorption of single-layer MoS2 deposited on PP and PDMS respectively calculated with energy-dependent $n_s$ and without an energy dependency. The absorption spectra of single-layer MoS$_2$ are almost identical with and without considering the energy dependence of the refractive index of the substrate, both in the case of PDMS and in the case of PP.

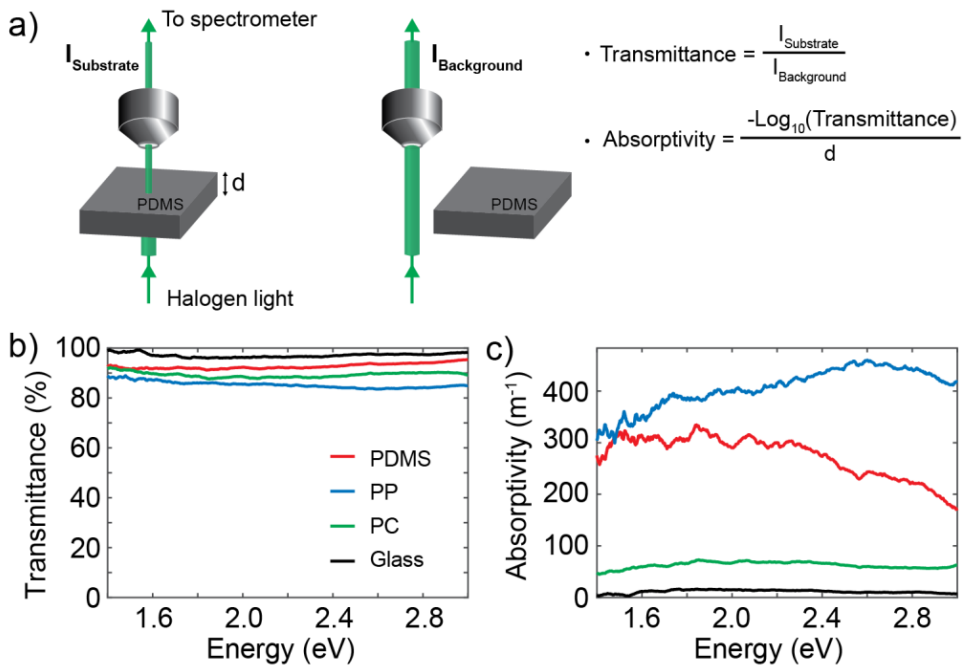

**Figure SI7**: a) Schematic drawing of the set-up configuration used to perform transmittance experiments. b-c) Experimental transmittance (b) and absorptivity (c) of four transparent substrates calculated from the formula in panel (a). The measured thickness is 110 µm for PDMS, 170 µm for PP, 790 µm for PC and 1120 µm for glass.

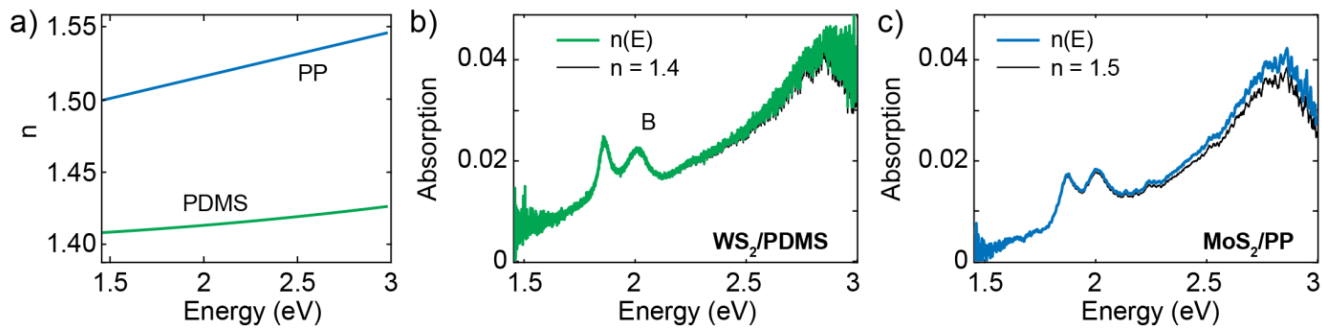

**Figure SI8:** a) Dispersion of refractive index of PP and PDMS with light wavelength [3,4]. b-c) Absorption spectra obtained with (colored) and without (black) considering the energy dependence of the refractive indexes of the two substrates using the equation (1) in the supplementary information.

### Section 4 – Optical images of single-layer TMDCs

Figure SI9 displays optical images of a $MoS_2$ flake with terraces of different thickness due to the different optical contrast; each image has been recorded in transmission mode and in reflection mode. This $MoS_2$ flake has been deposited onto the PDMS substrate by mechanical exfoliation with Nitto SPV 224 tape (Figure SI9a) and deterministically transferred onto a PP substrate (Figure SI9b). The flake was then subjected to three cycles of heating/cooling in a biaxial strain experiment and Figure SI9c displays the microscope pictures of the flake recorded after the experiment. Notice that the flake after the application of strain has ruptured at various points. Figure SI10 shows the flakes used for the differential reflectance measurements as a function of the substrate strain presented in Figure 4 of the main text. The single-layer regions of these flakes, which have been probed in the experiments, have lateral dimensions larger than 10 μm.

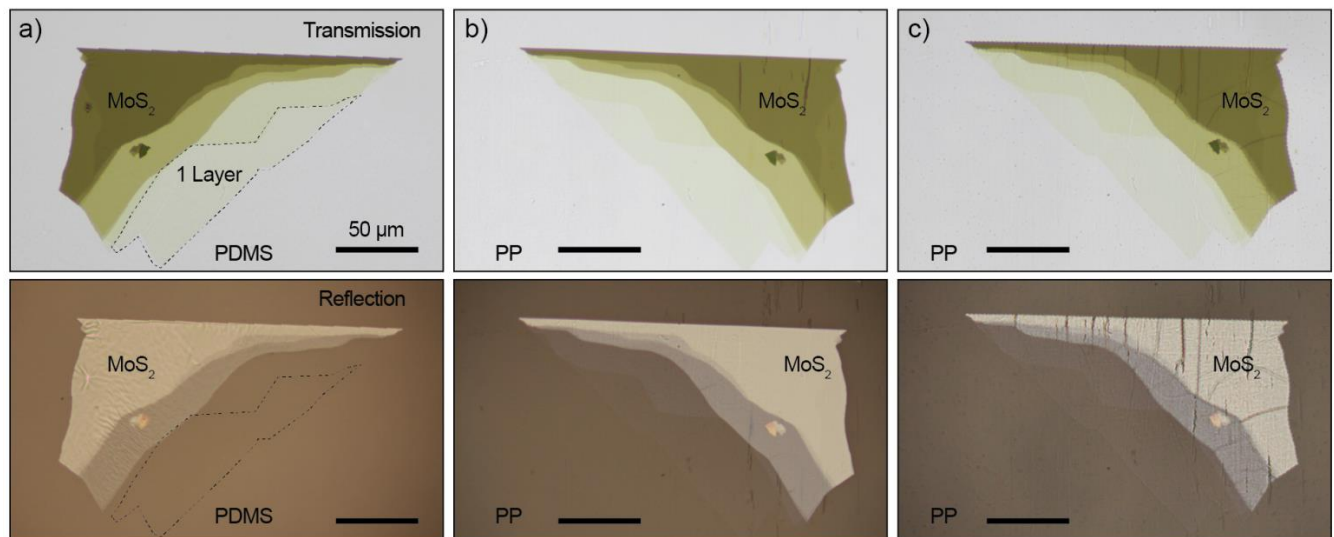

**Figure SI9**: a) Optical images of a $MoS_2$ flake recorded in transmission (top) and in reflection (bottom) deposited on a PDMS substrate. The dashed line is a guide for the eye and highlights the single-layer region of the $MoS_2$ flake. b) Optical images of the same $MoS_2$ flake depicted in (a) after deterministic transfer to a PP substrate. c) Same as (b) after three cycles of heating/warming in the biaxial strain experiment.

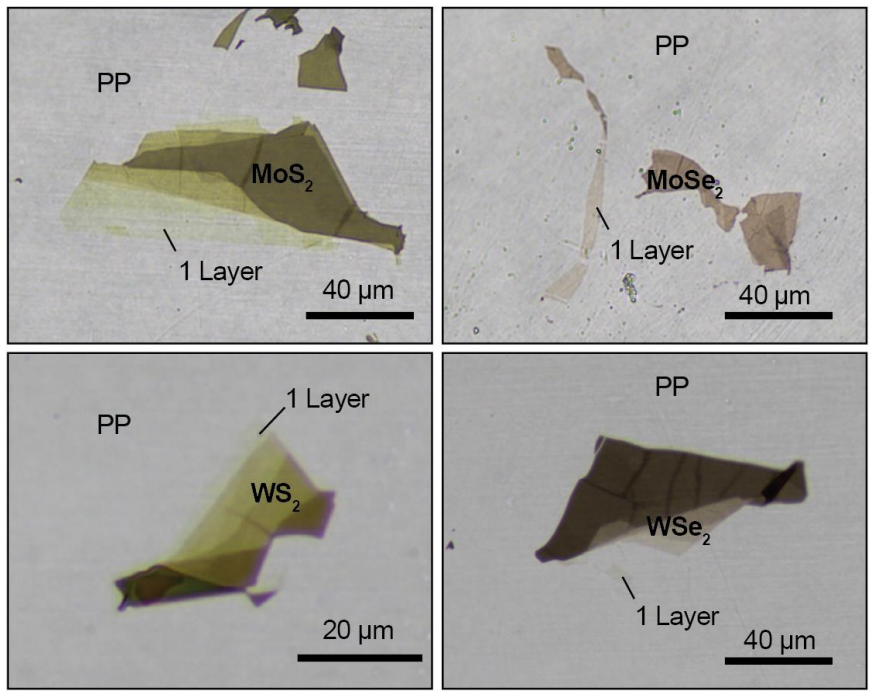

**Figure SI10**: Optical images of flakes of the four TMDCs studied recorded in transmission illumination mode deposited on a PP substrate.

S**ection 5 – Biaxial strain of single-layer TMDCs**

Figure SI11a displays the differential reflectance spectra of a single-layer $MoS_2$ flake deposited on a glass substrate recorded as function of the substrate temperature. Figure SI11b shows the position of the A and the B exciton, extracted from panel a, as a function of temperature. A linear fit to the data gives the intrinsic thermal shift of single-layer $MoS_2$.

Figure SI12a shows the differential reflectance spectra of $MoS_2$ single-layer flakes on polycarbonate (PC) substrate as a function of substrate temperature. Figure S12b shows the position of the A and the B exciton, extracted from panel a, as a function of temperature.

Figure SI13 shows differential reflectance spectra of single-layer flakes of the four TMDCs studied deposited on PDMS substrates. The spectra have been recorded as a function of the substrate temperature and from each spectrum we extracted the position of the various excitonic peaks. Table SI2 collects the gauge factor of each exciton for the four materials deposited on PDMS. Figure SI14 compares the differential reflectance spectrum of single-layer $MoS_2$ deposited on the three different substrates recorded at room temperature and zero strain. All the three spectra exhibit two maxima due to A and B excitons, whose position and width depend on the substrate. The difference between the energy of the A and B excitons is mostly due to the spin-orbit splitting of the valence band of these 2D TMDCs as schematized in Fig. SI15b. We find an excellent agreement between the predicted spin-orbit coupling energies and the experimental values as shown in Fig. SI15a.

In order to test the reproducibility of the strain application we performed many consecutive heating cooling cycles for single-layer MoS₂ deposited on PP and on PDMS. Figure SI16a shows the position of the A exciton peak measured on MoS₂ on PP during 6 consecutive cycles of warming/cooling the substrate between 25 °C and 95 °C. Over time the measurement appears reproducible giving similar values of the A exciton peak shift in each cycle. The same is valid for MoS₂ on PDMS shown in Fig. SI16b

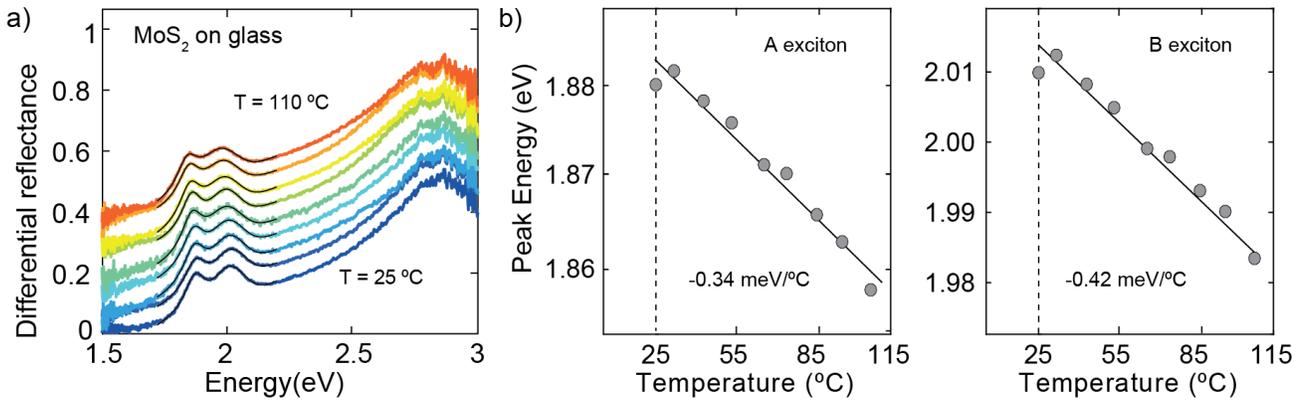

**Figure SI11**: a) Differential reflectance spectra of a single-layer MoS₂ flake deposited on glass as a function of the temperature. b) Energy of the A and B excitons of single-layer MoS₂ on glass. The solid black lines are linear fit to the data.

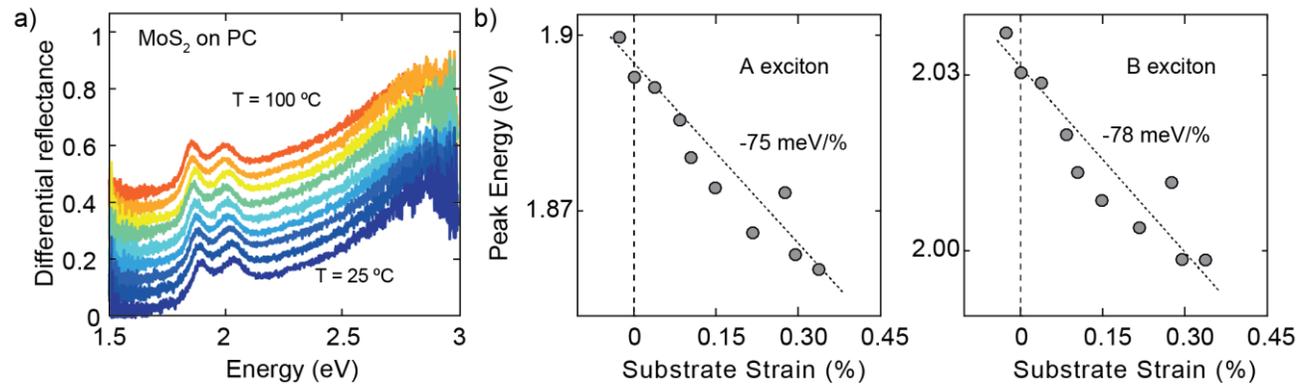

**Figure SI12**: a) Differential reflectance spectra of a single-layer MoS₂ flake deposited on polycarbonate (PC) as a function of the substrate expansion ($\alpha_{PC} \approx 0.45 \cdot 10^{-4}$ °C⁻¹). b) Energy of the A and B excitons of single-layer MoS₂ on PC. The black lines are linear fit to the data.

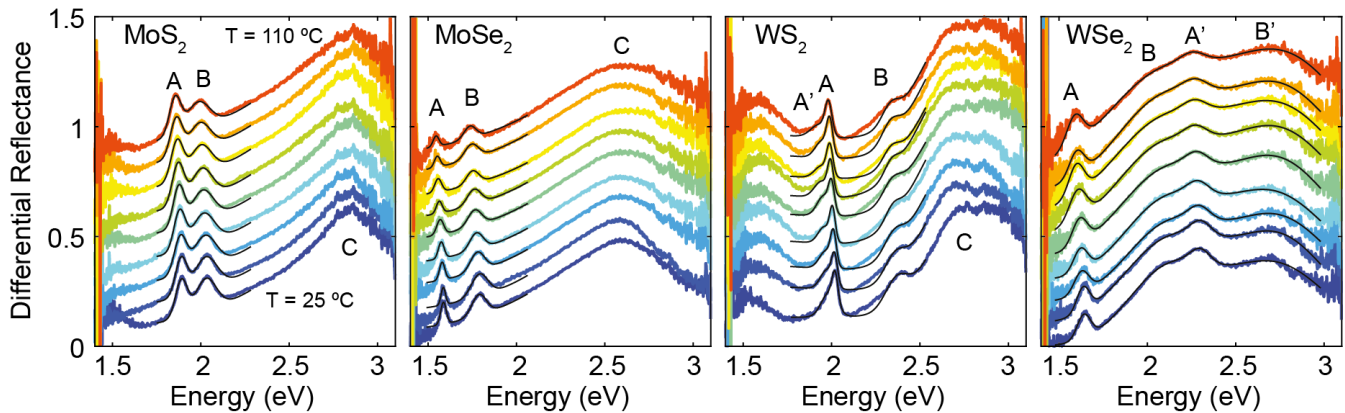

**Figure SI13**: Differential reflectance spectra of the four TMDCs investigated in this paper deposited on PDMS recorded as a function of the substrate temperature.

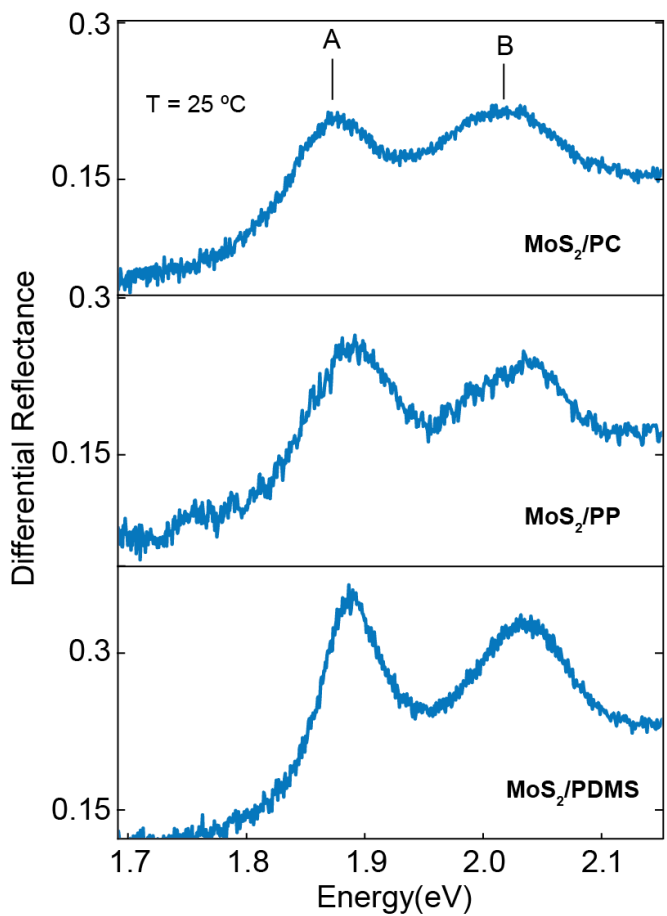

**Figure SI14**: a) Differential reflectance spectra of a single-layer MoS$_2$ flake deposited on PC (top), PP (middle) and PDMS (bottom) taken at room temperature.

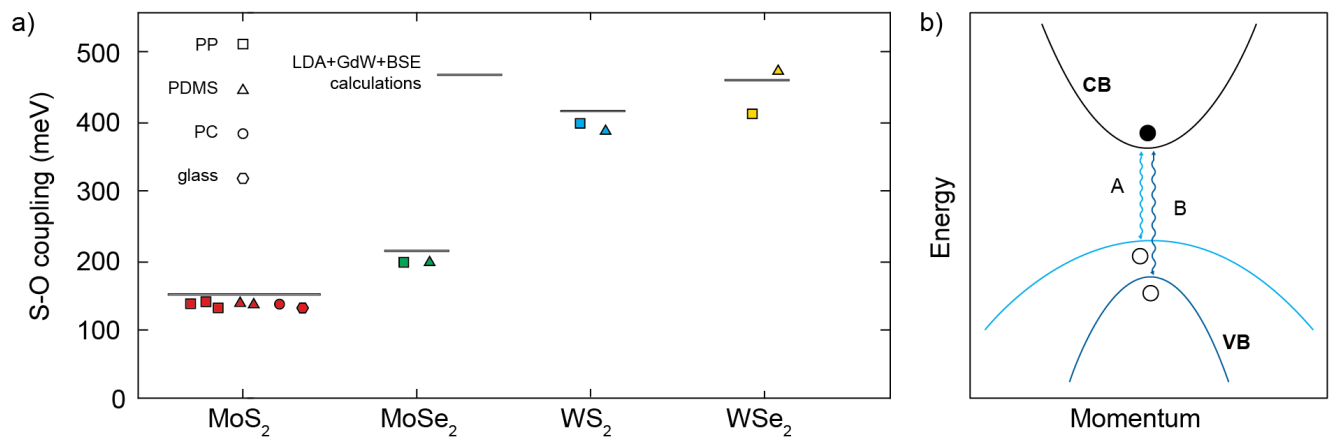

**Figure SI15**: a) Difference between the A and B exciton peaks energy for the different TMDCs studied in this manuscript as a function of strain. b) Schematic depiction of the band diagram of the single-layer TMDCs with indicated A and B excitons located at the direct transition point. The difference between the A and B excitons is approximately the spin-orbit splitting of the valence band of the TMDCs monolayer.

| Exciton | $MoS_2$/PDMS | $MoSe_2$/PDMS | $WS_2$/PDMS | $WSe_2$/PDMS |
|---|---|---|---|---|
| A | -13.3 meV/% | -15.3 meV/% | -13.0 meV/% | -18.4 meV/% |
| B | -15.8 meV/% | -21.4 meV/% | -11.7 meV/% | -19.5 meV/% |
| A' | | | -12.4 meV/% | -15.1 meV/% |
| B' | | | | 13.4 meV/% |

**Table SI2**: Gauge factor for excitons A (A') and B (B') extracted from applying biaxial strain on the different two-dimensional flakes deposited on PDMS. $WS_2$ and $WSe_2$ show additional excitonic features in their spectra called A' and B'. In the case of $WS_2$ the feature A' is due to the negatively charged A exciton (A' is a so-called trion), while $WSe_2$ displays two additional peaks, A' and B', which are due to higher-energy excitonic states.

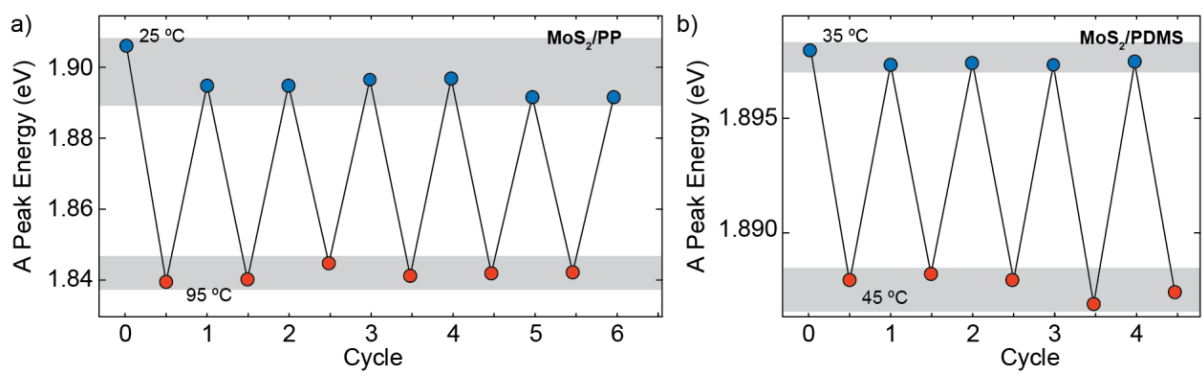

**Figure SI16**: a) Position of the A exciton of single-layer $MoS_2$ deposited on PP extracted from differential reflectance spectra measured at a temperature of 25 °C and 95 °C during consecutive heating/cooling cycles. The grey bands indicate the range explored in the values of A at low and high temperature. b) Same as (a) for single-layer $MoS_2$ on PDMS.

**Section 6 – Finite element analysis of biaxial strain transfer**

To understand the role of the substrate in the transfer of strain to single-layer TMDCs deposited on top, we performed a three-dimensional axisymmetric finite element analysis (FEA). The model shown in Fig. SI17a consists of single-layer MoS$_2$ with a thickness of 0.7 nm and Young's modulus $E_{MoS2}$ = 350 GPa, placed on PDMS substrate with a thickness of 100 µm. The FE model mesh was determined through a series of convergence studies. The interface between the MoS$_2$ flake and the substrate is modelled using perfect bonding. The calculations were performed using the commercial FE software COMSOL Multiphysics (version 5.2). In each step of the simulation we let the substrate expand thanks to thermal expansion and we extract the total expansion induced in the MoS$_2$ flake. Figure SI17b shows the amount of strain transferred as a function of the Young's modulus of the substrate. The results indicate that substrates with high Young's modulus ($E_{sub}$ > 1 GPa) can transfer a substantial amount of strain during tensile experiments. For example PP can transfer approximately 75% of the strain while PDMS, whose Young's modulus is $E_{PDMS} \approx 10^{-3}$ GPa, can only transfer 1% of the tensile strain to MoS$_2$. Figure SI18 shows the transferred strain for two different values of the Young's modulus of the flake.

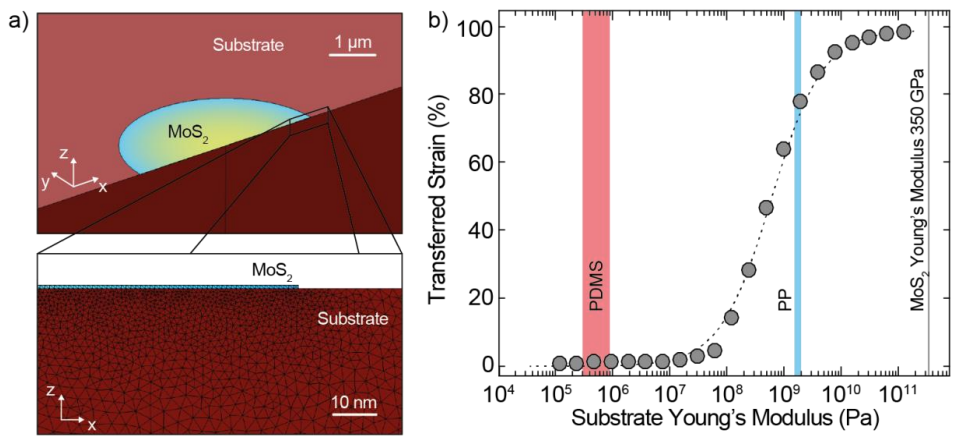

**Figure SI17**: a) Finite element calculation of a biaxial strain test sample consisting of a 100-µm-thick substrate and single-layer MoS$_2$ (0.7 nm thickness) on the substrate. b) Maximum transferred strain in MoS$_2$ as a function of substrate's Young's Modulus.

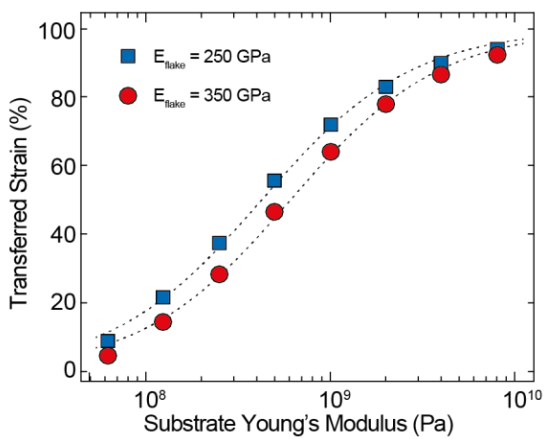

**Figure SI18**: Maximum transferred strain in a monolayer flake with Young's modulus of 250 GPa or 350 GPa, as a function of substrate's Young's Modulus.

**Section 7 – Calculating the absorption spectra under biaxial strain**

The first step of our *ab-initio* approach is a density-functional theory calculation (DFT) in the local density approximation (LDA). Here we employ norm-conserving pseudopotentials and a basis sets of three shells of localized Gaussian orbitals per atom for all four materials. The shells have s, p, d, and s* symmetry with decay constants between 0.13 $a_B^{-2}$ and 2.5 $a_B^{-2}$. For integrations in the reciprocal space, a k-mesh of $10 \times 10 \times 1$ is used. A large interlayer distance of 45 Å is employed to suppress interlayer interaction. For all following calculations the optimized structure is used, with forces smaller than $10^{-4}\ Ry/a_B$. We obtain theoretical lattice constants of (3.16, 3.30, 3.15, 3.29)Å for (MoS$_2$, MoSe$_2$, WS$_2$, WSe$_2$) which agree very well with the experimental values of (3.16 [5], 3.30 [5], 3.16 [6], 3.28 [7]) Å.

Quasiparticle calculations are carried out within the LDA+*GdW* approximation using an auxiliary plane wave basis with a 2.5 Ry energy cutoff (205 plane waves). The left panel of Figure SI19 shows the fast convergence behavior of the direct quasiparticle gap at the K point with respect to the energy cutoff. For the electronic structure (e.g. the gap), interlayer interactions between different supercells are circumvented by calculating several interlayer distances $L$ (up to 60 Å) and interpolating to $L \to \infty$. Spin-orbit interaction is fully included in the DFT and the quasiparticle calculations.

Absorption spectra are obtained by solving the Bethe-Salpeter equation (BSE). Here, we employ identical meshes for the k-points where the quasiparticle corrections and the electron-hole interactions are calculated, omitting the need of an interpolation scheme ($30 \times 30 \times 1$ k points). Note that in our approach the excitation energies converge extremely fast with respect to the grid, as can be seen in Figure SI19 in the right panel. Six valence and four conduction bands are included in the setup of the BSE Hamiltonian.

All calculations are carried out using a code written by ourselves [8]. Figure SI20 displays the absorption spectra of the four single-layer TMDCs studied in this paper.

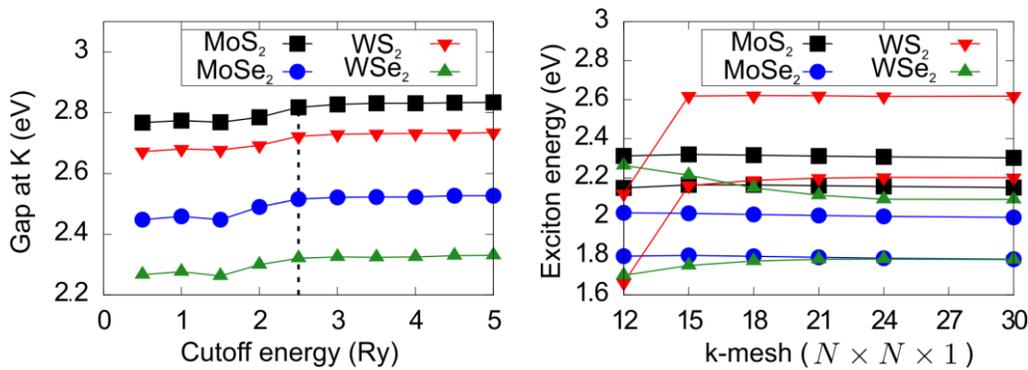

**Figure SI19**: Left panel: Convergence of the direct quasiparticle gap at the K point with respect to the energy cutoff of the plane wave basis for the given TMDCs. The interlayer distance is fixed at 45Å and the k-mesh is fixed at $20 \times 20 \times 1$. The solid lines are a guide to the eye and the dashed line is the cutoff applied. All materials show similarly fast convergence behavior. Right panel: Convergence of the A and B exciton with respect to the k-grid applied in the BSE. For each material, the lower energy excitation represents the A exciton.

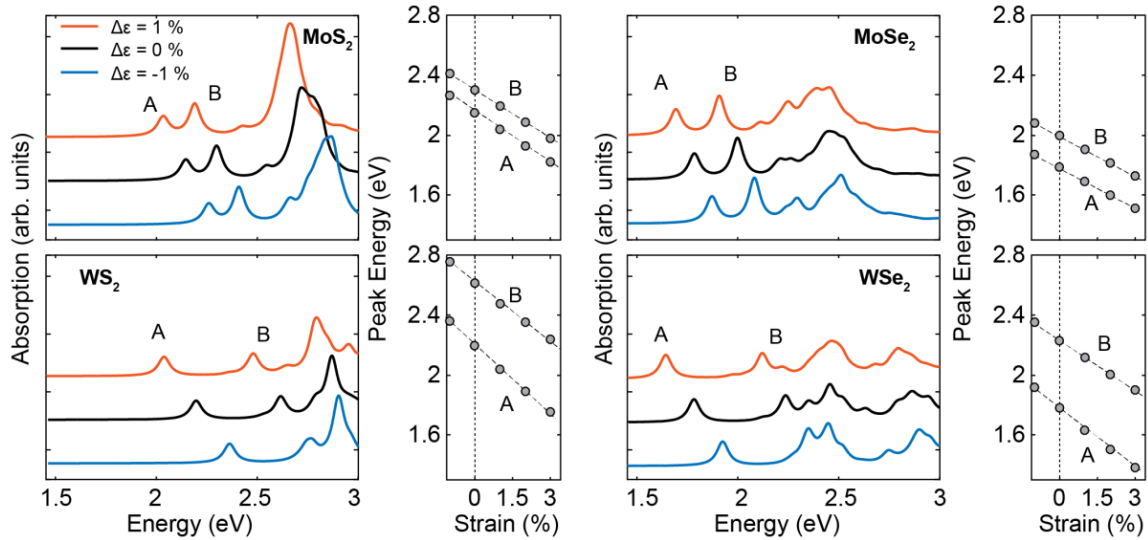

**Figure SI20**: Calculated BSE absorption spectra under biaxial strain. An artificial broadening of 0.035 eV is used in the BSE and the spectra are vertically shifted for improved visibility. Energies of the A and B excitons are extracted from the ab-initio calculation.